\documentclass[fleqn,10pt]{wlscirep}
\usepackage{amssymb}
\usepackage{amsmath}
\usepackage{graphicx}
\usepackage{epsfig}

\title{Stimulated photon emission and two-photon Raman scattering in a
coupled-cavity QED system}

\author[1]{C. Li}
\author[1,*]{Z. Song}
\affil[1]{School of Physics, Nankai University, Tianjin 300071, China}

\affil[*]{songtc@nankai.edu.cn}

\begin{abstract}
We study the scattering problem of photon and polariton in a
one-dimensional coupled-cavity system. Analytical approximate analysis and
numerical simulation show that a photon can stimulate the photon emission
from a polariton through polariton-photon collisions. This observation opens
the possibility of photon-stimulated transition from insulating to radiative
phase in a coupled-cavity QED system. Inversely, we also find that a
polariton can be generated by a two-photon Raman scattering process. This
paves the way towards single photon storage by the aid of atom-cavity
interaction.
\end{abstract}

\begin{document}
\maketitle


\flushbottom

%
%


\section*{Introduction}

A coupled-cavity QED system provides a promising platform to study novel
quantum phenomena, since it combines two or more distinct quantum
components, exhibiting features not seen in these individual systems. The
discrete spatial mode of the photon in a coupled-cavity array and its nonlinear
coupling to atom make the possible applications both in quantum information
processing \cite{Kimble} and quantum simulation \cite{Nori}.\textbf{\ }The
seminal papers \cite{Hartmann, Angelakis, greentree} proposed the use of the
system to create strongly correlated many-body models. It has predicted the
quantum phase transition from Mott insulator phase to superfluid phase \cite%
{greentree,Huo}. This scenario is constructed under the assumption that
there is no extra photon leaking into the system. The stability of an
insulating phase bases on the fact that the polariton states in a cavity QED
system are eigenstates, i.e., spontaneous photon emission is forbidden. This
situation may change if a photon can stimulate the photon emission from a
polariton. In contrast with quantum phase transition induced by varying
system parameters, such as atom-cavity coupling strength, stimulated photon
emission from polaritons can also trigger the transition between insulating
and radiative phases. It is interesting and important to investigate the
photon-photon and photon-polariton scattering processes.\textbf{\ }Many
efforts related to few-body dynamics mainly focused on multi-photon
transports through coupled-cavity QED systems \cite{Moorad, Roy, Liao, Roy1,
Shi1, Eden, Ben, Shi2, Roy2, Roy3, Eden1, Xu, Xun}\textbf{, }while a few
works dealt with the formation of bound state \cite{Shi, Longo, Zheng}. So
far, what happens when a photon collides with a polariton is still an open
question.

In this paper, we study the scattering problem of an incident photon by a
polariton in a one-dimensional coupled-cavity QED system. Analytical
approximate analysis and numerical simulation reveal several dynamical
features. We find that a photon can stimulate the photon emission from a
polariton, which induces the amplification of the photon population in a
multi-polariton system. After a chain reaction, incident photons can
stimulate the transition from insulating to radiative phases in the system
with low doped cavity density. We also investigate the inverse process of
stimulated photon emission from a polariton. We will show that a polariton
can be generated by a two-photon Raman scattering process, which has been
studied for the atoms found in nature\textbf{\ }\cite{Schrey, Puentes, Kim}.
Moreover, it has been shown that an atom-cavity system can behave as a
quantum switch for the coherent transport of a single photon \cite{zhou1}.
Considering a two-excitation problem, we find that a single-photon
transmission through a quantum switch is affected significantly by a
polariton that resides at it.

This paper is organized as follows. At first, we present the model and
single-excitation polaritonic states. Then, we propose an effective
Hamiltonian to analyze the possibility of photon emission from two aspects.
Numerical simulations for two-particle collision processes are showed later.
Finally, we give a summary and discussion.

\section*{Results}


\subsection*{Model and polariton}

We consider a one-dimensional coupled-cavity system with a two-level atom,
which is embedded in the center of cavity array. The Hamiltonian can be
written as
\begin{equation}
H=-\kappa \sum_{\left\vert l\right\vert =0}^{N}a_{l}^{\dag }a_{l+1}+\lambda
a_{0}^{\dag }\left\vert g\right\rangle \left\langle e\right\vert +\text{%
\textrm{H.c.}},  \label{H}
\end{equation}%
where $\lambda $ represents atom-cavity coupling strength and $\kappa $ is
the photon hopping strength for the tunneling between adjacent cavities.
Here, $\left\vert g\right\rangle $ ($\left\vert e\right\rangle $) denotes
the ground (excited) state of the qubit with $\sigma ^{z}\left\vert
e\right\rangle =\left\vert e\right\rangle $ and $\sigma ^{z}\left\vert
g\right\rangle =-\left\vert g\right\rangle $, $a_{l}$ ($a_{l}^{\dag }$)
annihilates (creates) a photon at the $l$th cavity. Obviously the total
excitation number, $\mathcal{\hat{N}}\mathcal{=}\sum_{\left\vert
l\right\vert =0}^{N}a_{l}^{\dag }a_{l}+\sigma ^{z}+\frac{1}{2}$, is a
conserved quantity for the Hamiltonian $H$, i.e., $[H,\mathcal{\hat{N}}]=0$.

The coupled-cavity array can be considered as a one-dimensional waveguide,
while the two-level atom can act as a quantum switch to control the
single-photon transmission \cite{zhou1}. To demonstrate this point, we
rewrite the Hamiltonian in the form

\begin{equation}
\overline{H}=-2\kappa \sum_{k}\cos ka_{k}^{\dag }a_{k}+\frac{\lambda }{\sqrt{%
2N}}\sum_{k}\left( a_{k}^{\dag }\left\vert g\right\rangle \left\langle
e\right\vert +\text{\textrm{H.c.}}\right),  \label{H_k}
\end{equation}%
where%
\begin{eqnarray}
a_{k}^{\dag } &=&\frac{1}{\sqrt{2N}}\sum_{\left\vert l\right\vert
=0}^{N}e^{ikl}a_{l}^{\dag }, \\
a_{l}^{\dag } &=&\frac{1}{\sqrt{2N}}\sum_{k}e^{-ikl}a_{k}^{\dag }.
\end{eqnarray}%
It indicates that the atom couples to photons of all modes $k\in \left[ -\pi
,\pi \right] $. In the $\mathcal{N}=1$\ subspace, atom can be regarded as a
stationary scattering center. All the dynamics can be treated in the context
of single-particle scattering method, which has been well studied \cite%
{zhou1}.

A comprehensive understanding for the dynamics involving the sector with $%
\mathcal{N}>1$ is necessary to both theoretical explorations and practical
applications. Intuitively, the state of the atom ($\left\vert e\right\rangle
$\ or $\left\vert g\right\rangle $) should affect the interaction between
the atom and a photon. In experiments, the practical processes may concern
two or more photons, which obviously affect on the function of the quantum
switch. On the other hand, the stability of an insulating phase may be
spoiled by the background photons from environment. In this paper, we study
the scattering problem in the $\mathcal{N}=2$\ sector, focusing on the
effect of the nonlinearity arising from the atom. The investigation has two
aspects: First, we study the photon scattering from a polariton. Secondly,
we consider the collision of two photons under the atom-cavity nonlinear
interaction.

We start our investigation with the solution of single-particle bound and
scattering states. In the invariant subspace with $\mathcal{N=}1$, exact
solution shows that there are two bound states, termed as single-excitation
polaritonic states, being the mixture of photonic and atomic excitations.
From the Method, these polaritonic states are obtained by Bethe Ansatz
method as the form%
\begin{equation}
\left\vert \phi ^{\pm }\right\rangle =\pm \frac{2\kappa }{\lambda \sqrt{%
\Omega }}\sinh \beta \left\vert e\right\rangle \left\vert 0\right\rangle
+\sum\limits_{\left\vert l\right\vert =0}\frac{\left( \mp 1\right) ^{l}}{%
\sqrt{\Omega }}e^{-\beta l}a_{l}^{\dagger }\left\vert g\right\rangle
\left\vert 0\right\rangle ,
\end{equation}%
where the normalization factor is%
\begin{equation}
\Omega =\left( \frac{2\kappa }{\lambda }\right) ^{2}\sinh ^{2}\beta +\coth
\beta
\end{equation}%
and%
\begin{equation}
\left\vert 0\right\rangle =\prod_{\left\vert l\right\vert =0}\left\vert
0\right\rangle _{l},a_{l}\left\vert 0\right\rangle _{l}=0
\end{equation}%
The corresponding energy is

\begin{equation}
\varepsilon _{\pm }=\pm 2\kappa \cosh \beta ,
\end{equation}%
where the positive number $\beta $ determines the extension of bound states
around the doped cavity, obeys the equation%
\begin{equation}
e^{2\beta }=\sqrt{\left( \lambda /\sqrt{2}\kappa \right) ^{4}+1}+\left(
\lambda /\sqrt{2}\kappa \right) ^{2}.
\end{equation}%
We can see that $\beta $\ has nonzero solutions for nonzero $\lambda $,
indicating the existence of nontrivial bound states.

On the other hand, the derivation in Method shows that the solution of
scattering states $\left\vert \phi ^{k}\right\rangle $\ with energy $%
\varepsilon _{k}=-2\kappa \cos k$ has the form

\begin{equation}
\left\vert \phi ^{k}\right\rangle =\frac{1}{\sqrt{\Lambda _{k}}}\{\left\vert
e\right\rangle \left\vert 0\right\rangle +\frac{\epsilon _{k}}{\lambda }%
a_{0}^{\dagger }\left\vert g\right\rangle \left\vert 0\right\rangle +\frac{1%
}{4i\kappa \lambda \sin k}\sum\limits_{l\neq 0,\sigma =\pm }\varsigma
_{\sigma }e^{i\sigma k\left\vert l\right\vert }a_{l}^{\dag }\left\vert
g\right\rangle \left\vert 0\right\rangle \},  \label{scattering state}
\end{equation}%
where $\Lambda _{k}$\ is the normalization factor and%
\begin{equation}
\varsigma _{\pm }=\pm \left[ \left( \lambda ^{2}-\epsilon _{k}^{2}\right)
2\kappa \epsilon _{k}e^{\mp ik}\mp \left( \lambda ^{2}-\epsilon
_{k}^{2}\right) \right] .
\end{equation}

\begin{figure}[tbp]
\centering
\includegraphics[ bb=47 427 563 730, width=0.45\textwidth, clip]{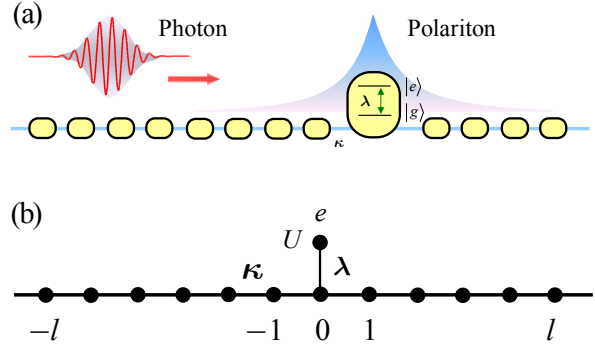}
\caption{(Color online) (a) Schematic configuration for the coherent
collision of polariton and photon. An array of coupled single-mode cavities,
where the central cavity is coupled to a two-level atom. Initially a
polariton is located at the center, while a photon wave packet is moving
from the left to collide with the polariton. (b) Schematic illustration for
the equivalent description of the hybrid system. The excited state of the
atom can be treated as a side-coupling site with infinite on-site repulsion.}
\label{fig1}
\end{figure}

We can see that a polariton is a local eigen state of the system, which is
stable and cannot emit a photon in the $\mathcal{N=}1$\ subspace. The aim of
this work is considering the effects of photon-photon and photon-polariton\
collisions. Our strategy is sketched in Fig. \ref{fig1}(a). In the invariant
subspace with $\mathcal{N=}2$, a two-excitation state can be a direct
product of a local photon and a polariton states, which are well separated
in real space. As long as time evolution, two local particles are
overlapped. The nonlinear effect induces the interaction between the photon
and polariton. After a relaxation time, the free photons spread out from the
central cavity, only the polaritons are left, being stationary at the
center. In the case of the ultimate polaritonic probability being less than $%
1$, (or the escaped photon number larger than $1$) we can conclude that the
stimulated photon emission occurs during the process. We will show that this
behavior becomes crucial when we study the stability of a macroscopic
insulating phase, and the efficiency of a quantum switch in a waveguide.\ In
the following section, we will analyze the possibility of photon emission
from two aspects.

\subsection*{Effective description}

In this section, we present an analytical analysis on the effects of
photon-photon and photon-polariton\ collisions. This will be based on an
effective description of the original Hamiltonian $H$ or $\overline{H}$. We
extend the Hilbert space by introducing the auxiliary\ photon state\ $\left(
a_{e}^{\dagger }\right) ^{n}\left\vert 0\right\rangle _{e}$,\ where $%
a_{e}^{\dagger }$\ is the creation operator of a photon at site $e$ and $%
\left\vert 0\right\rangle _{e}$ is the corresponding vacuum state. The qubit
state $\left\vert e\right\rangle $ is replaced by $a_{e}^{\dagger
}\left\vert 0\right\rangle _{e}$. We rewrite the original Hamiltonians $H$
and $\overline{H}$ as the Hubbard models%
\begin{equation}
H_{\mathrm{eq}}=-\kappa \sum_{\left\vert l\right\vert =0}^{N}a_{l}^{\dag
}a_{l+1}+\lambda a_{0}^{\dag }a_{e}+\text{\textrm{H.c.}}+\frac{U}{2}%
a_{e}^{\dagger }a_{e}\left( 1-a_{e}^{\dagger }a_{e}\right) ,  \label{H_eq1}
\end{equation}%
and%
\begin{equation}
\overline{H}_{\mathrm{eq}}=-2\kappa \sum_{k}\cos ka_{k}^{\dag }a_{k}+\frac{%
\lambda }{\sqrt{2N}}\sum_{k}\left( a_{k}^{\dag }\left\vert g\right\rangle
\left\langle e\right\vert +\text{\textrm{H.c.}}\right) +\frac{U}{2}%
a_{e}^{\dagger }a_{e}\left( 1-a_{e}^{\dagger }a_{e}\right) .  \label{H_eq2}
\end{equation}%
We note that the state $\left( a_{e}^{\dagger }\right) ^{n}\left\vert
0\right\rangle _{e}$ with $n>1$ will be ruled out as $U\rightarrow \infty $,
Hamiltonians$\ H_{\mathrm{eq}}$\ and $\overline{H}_{\mathrm{eq}}$\ being
equivalent to $H$ and $\overline{H}$, respectively. Correspondingly, we have
$[H_{\mathrm{eq}},\mathcal{\hat{N}}_{\mathrm{eq}}]=[\overline{H}_{\mathrm{eq}%
},\mathcal{\hat{N}}_{\mathrm{eq}}]=0$\ by defining $\mathcal{\hat{N}}_{%
\mathrm{eq}}\mathcal{=}\sum_{\left\vert l\right\vert =0}^{N}a_{l}^{\dag
}a_{l}$ $+a_{e}^{\dagger }a_{e}$. We will see that this equivalence can be
true for a large magnitude $U\sim 10$. Next, we will perform our analysis
from two aspects: $k$ space and real space.

\subsection*{Coupled equations in $k$ space}

\begin{figure}[tbp]
\centering
\includegraphics[ bb=36 457 547 759, width=0.45\textwidth, clip]{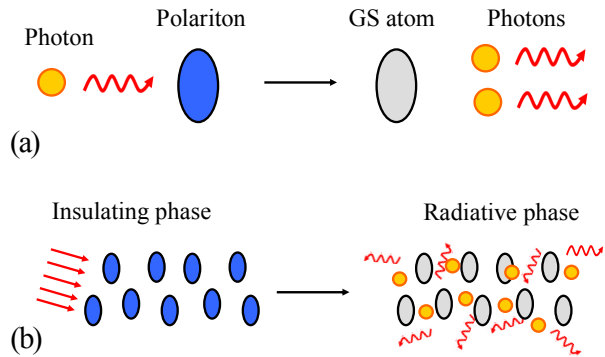}
\caption{(Color online) Polariton-photon transition in a coupled-cavity
array coupled to a two-level atomic system. (a) When the collision between a
photon and polariton occurs, the total photon probability cannot be
preserved. The gain of photons indicates the stimulated photon emission.\
The blue (gray) color represents the polaritonic (atomic ground) state. (b)
The insulating-radiative phase transition. A multi-polaritonic insulating
state can collapse to a radiative state by an external field radiation.}
\label{fig2}
\end{figure}

\begin{figure}[tbp]
\centering
\includegraphics[ bb=35 431 562 760, width=0.45\textwidth, clip]{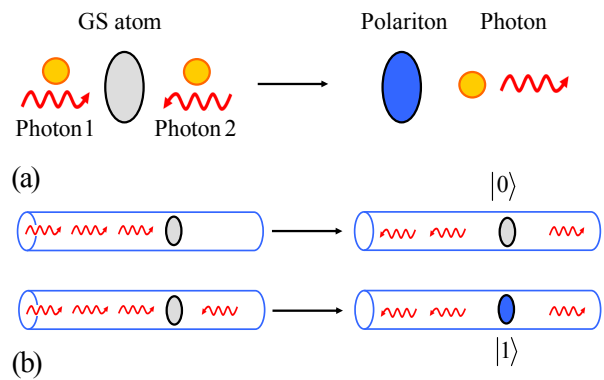}
\caption{(Color online) The two-photon Raman transition in a coupled-cavity
array coupled to a two-level atomic system. (a) When the collision between
two photons from opposite directions occurs at the cavity with an atom, the
total photon probability cannot be preserved. The loss of photons indicates
the two-photon Raman transition.\ The blue (gray) color represents the
polaritonic (atomic ground) state. (b) Single-photon storage by the aid of
single photons train from the opposite side. Any single photons cannot be
stored in the atom when they transmit unidirectionally. It can be achieved
by an incident single photons from the opposite side.}
\label{fig3}
\end{figure}

First of all, we would like to point out that the eigenstates of the
Hamiltionians $H$ and $\overline{H}$ in Eqs. (\ref{H}) and (\ref{H_k}) are
still the eigenstates of $H_{\mathrm{eq}}$\ and $\overline{H}_{\mathrm{eq}}$
by taking $\left\vert e\right\rangle \rightarrow a_{e}^{\dagger }\left\vert
0\right\rangle _{e}$. Now we consider the case in two-particle subspace. The
basis set for two-particle invariant subspace can be constructed by the
single-particle eigen states $\left\vert \phi ^{\pm }\right\rangle $\ and $%
\left\vert \phi ^{k}\right\rangle $. We concern the complete basis set with
even parity, which can be classified into four groups%
\begin{eqnarray}
\left\{ \left\vert 1,\sigma ,k\right\rangle \right\} &:&\left\vert \phi
^{\sigma }\right\rangle \left\vert \phi ^{k}\right\rangle , \\
\left\{ \left\vert 2,k,k^{\prime }\right\rangle \right\} &:&\left\vert \phi
^{k}\right\rangle \left\vert \phi ^{k^{\prime }}\right\rangle , \\
\left\{ \left\vert 3,\sigma ,\sigma ^{\prime }\right\rangle \right\}
&:&\left\vert \phi ^{\sigma }\right\rangle \left\vert \phi ^{\sigma ^{\prime
}}\right\rangle .
\end{eqnarray}%
where $\sigma =\pm $. We note that state $\left\vert \phi ^{k}\right\rangle
\left\vert \phi ^{k^{\prime }}\right\rangle $\ is automatically the
eigenstate of $H$ with eigen energy $\varepsilon _{k}+\varepsilon
_{k^{\prime }}$. And states $\left\vert \phi ^{\sigma }\right\rangle
\left\vert \phi ^{\sigma ^{\prime }}\right\rangle $\ will be ruled out as $%
U\rightarrow \infty $. Then basis sets $\left\{ \left\vert 1,\sigma
,k\right\rangle \right\} $\ and $\left\{ \left\vert 2,k,k^{\prime
}\right\rangle \right\} $\ can further construct an invariant subspace
approximately. In this sense, the solution of the Schrodinger equation
\begin{equation}
i\frac{\partial }{\partial t}\left\vert \psi \left( t\right) \right\rangle
=H\left\vert \psi \left( t\right) \right\rangle ,  \label{seq}
\end{equation}%
has the form%
\begin{equation}
\left\vert \psi \left( t\right) \right\rangle =\sum_{k,\sigma =\pm
}C_{1,\sigma ,k}\left( t\right) \left\vert \phi ^{\sigma }\right\rangle
\left\vert \phi ^{k}\right\rangle +\sum_{k,k^{\prime }}C_{2,k,k^{\prime
}}\left( t\right) \left\vert \phi ^{k}\right\rangle \left\vert \phi
^{k^{\prime }}\right\rangle ,
\end{equation}%
where coefficients $C_{1,\sigma ,k}\left( t\right) $ and $C_{2,k,k^{\prime
}}\left( t\right) $ describe the two-particle dynamics and satisfy the
coupled differential equations\
\begin{eqnarray}
i\frac{\partial }{\partial t}\mathbf{C}_{1}\left( t\right) &=&M_{11}\mathbf{C%
}_{1}\left( t\right) +M_{12}\mathbf{C}_{2}\left( t\right) ,
\label{coupled eqs} \\
i\frac{\partial }{\partial t}\mathbf{C}_{2}\left( t\right) &=&M_{22}\mathbf{C%
}_{2}\left( t\right) +M_{21}\mathbf{C}_{1}\left( t\right) .
\end{eqnarray}%
Here the column vectors $\mathbf{C}_{1}\left( t\right) =\left\{ C_{1,\sigma
,k}\left( t\right) \right\} $\ and $\mathbf{C}_{2}\left( t\right) =\left\{
C_{2,k,k^{\prime }}\left( t\right) \right\} $, and the matrix%
\begin{equation}
\left[
\begin{array}{cc}
M_{11} & M_{12} \\
M_{21} & M_{22}%
\end{array}%
\right]
\end{equation}%
is a matrix representation of $H$ on the basis set $\left\{ \mathbf{C}%
_{1}\left( t\right) ,\mathbf{C}_{2}\left( t\right) \right\} $. Although we
cannot get an analytical solution of $\left\vert \psi \left( t\right)
\right\rangle $, we can conclude that the nontrivial solution $\left\vert
\psi \left( t\right) \right\rangle $\ should predict the following relations
in principle. We can always have nonzero $\mathbf{C}_{2}\left( t\right) $\
from initial condition $\mathbf{C}_{1}\left( 0\right) \neq 0$\ but $\mathbf{C%
}_{2}\left( 0\right) =0$, i.e.,%
\begin{equation}
\left\vert \phi ^{\sigma }\right\rangle \left\vert \phi ^{k}\right\rangle
\longrightarrow \left\vert \phi ^{k^{\prime \prime }}\right\rangle
\left\vert \phi ^{k^{\prime }}\right\rangle ,  \label{process1}
\end{equation}%
and inversely, nonzero $\mathbf{C}_{1}\left( t\right) $\ from initial
condition $\mathbf{C}_{2}\left( 0\right) \neq 0$\ but $\mathbf{C}_{1}\left(
0\right) =0$, i.e.,%
\begin{equation}
\left\vert \phi ^{k}\right\rangle \left\vert \phi ^{k^{\prime
}}\right\rangle \longrightarrow \left\vert \phi ^{\sigma }\right\rangle
\left\vert \phi ^{k^{\prime \prime }}\right\rangle .  \label{process2}
\end{equation}%
The former corresponds to the stimulated photon emission of the polariton,
while the latter corresponds to the polariton state generation by a
two-photon Raman scattering. The two processes are schematically illustrated
in Figs. \ref{fig2}(a) and \ref{fig3}(a).

\subsection*{Effective photon blockade}

In this section, we will demonstrate the process in Eq. (\ref{process1})
from an alternative way. One can consider the collision between an incident
photon and an initial bound state around the site $e$ in the system $H_{%
\mathrm{eq}}$. The obtained result should be close to that of the $H$
system. In this context, the photon-photon collision only occurs at site $e$%
. Then the impact of the incident photon on the bound photon can be
approximately regarded as a kicked potential on the $e$th site. In the
following, we will investigate the effect the potential works on the
dynamics of the bound photon.

We reduce the two-particle system of $H_{\mathrm{eq}}$ to a single-particle
system with the effective time-dependent Hamiltonian,

\begin{eqnarray}
H_{\mathrm{eff}}\left( t\right)  &=&H_{0}+V\left( t\right) ,  \label{H_eff}
\\
H_{0} &=&-\kappa \sum_{\left\vert l\right\vert =0}^{N}\left\vert
l\right\rangle \left\langle l+1\right\vert +\lambda \left\vert
0\right\rangle \left\langle e\right\vert +\text{\textrm{H.c.,}} \\
V\left( t\right)  &=&U_{0}\delta \left( t-\tau \right) \left\vert
e\right\rangle \left\langle e\right\vert .
\end{eqnarray}%
where $U_{0}$\ is the strength of the scattering and $\left\vert
e\right\rangle =a_{e}^{\dagger }\left\vert 0\right\rangle _{e}$, $\left\vert
l\right\rangle =a_{l}^{\dagger }\left\vert 0\right\rangle $ $\left( l=0,\pm
1,\pm 2,...\right) $ denotes the single-photon state. The initial state is
one of the bound states%
\begin{equation}
\left\vert \phi ^{\pm }\right\rangle =\frac{1}{\sqrt{\Omega }}[\pm \frac{%
2\kappa }{\lambda }\sinh \beta \left\vert e\right\rangle
+\sum\limits_{l>0}\left( \mp 1\right) ^{l}e^{-\beta l}\left\vert
l\right\rangle ].  \label{bound+/-}
\end{equation}%
After the impact of the kicked potential, $\left\vert \phi ^{\pm
}\right\rangle $ should probably jump to the scattering states $\left\vert
\phi ^{k}\right\rangle $. In the following, we demonstrate this point based
on time-dependent perturbation theory.

For small $U_{0}$, the transition probability amplitude from the initial
state\ $\left\vert \varphi ^{\mu }\right\rangle $\ at $t=0$\ to $\left\vert
\varphi ^{\nu }\right\rangle $\ $\left( \mu ,\nu =\pm \right) $\ at\ $t>\tau
$\ can express as%
\begin{eqnarray}
A_{\mu \nu } &=&\delta _{\mu \nu }-i\int_{0}^{t}\left\langle \phi ^{\nu
}\right\vert V\left( t^{\prime }\right) \left\vert \phi ^{\mu }\right\rangle
e^{-i\left( \varepsilon _{\mu }-\varepsilon _{\nu }\right) t^{\prime }}%
\mathrm{d}t^{\prime } \\
&&-\underset{\eta =k,\pm }{\sum }\int_{0}^{t}\mathrm{d}t^{\prime
}\int_{0}^{t^{\prime }}\mathrm{d}t^{^{\prime \prime }}e^{-i\left(
\varepsilon _{\eta }-\varepsilon _{\nu }\right) t^{\prime }}\left\langle
\phi ^{\nu }\right\vert V\left( t^{\prime }\right) \left\vert \phi ^{\eta
}\right\rangle  \notag \\
&&\times \left\langle \phi ^{\eta }\right\vert V\left( t^{\prime \prime
}\right) \left\vert \phi ^{\mu }\right\rangle e^{-i\left( \varepsilon _{\mu
}-\varepsilon _{\eta }\right) t^{\prime \prime }},  \notag
\end{eqnarray}%
up to second order according to the time-dependent perturbation theory.
Using the identity%
\begin{equation}
\left\langle \phi ^{\mu }\right\vert V\left( t\right) \left\vert \phi ^{\eta
}\right\rangle =\left( -1\right) ^{1+\delta _{\mu \nu }}\left\langle \phi
^{\eta }\right\vert V\left( t\right) \left\vert \phi ^{\nu }\right\rangle ,
\end{equation}%
and the completeness condition

\begin{equation}
\sum_{\eta =k,\pm }\langle e\left\vert \phi ^{\eta }\right\rangle
\left\langle \phi ^{\eta }\right\vert e\rangle =1,
\end{equation}%
we get the transition probability between two bound states%
\begin{eqnarray}
T_{+-} &=&\left\vert A_{+-}\right\vert ^{2}=U_{0}^{2}p^{2}\left(
1+U_{0}^{2}\right) , \\
T_{\pm \pm } &=&\left\vert A_{\pm \pm }\right\vert ^{2}=\left(
1-U_{0}^{2}p\right) ^{2}+\left( U_{0}p\right) ^{2},
\end{eqnarray}%
where%
\begin{equation}
p=\langle \phi ^{\mu }\left\vert e\right\rangle \left\langle e\right\vert
\phi ^{\nu }\rangle =\left( -1\right) ^{1+\delta _{\mu \nu }}\frac{4\kappa
^{2}}{\lambda ^{2}\Omega }\sinh ^{2}\beta .
\end{equation}%
The crucial conclusion is that the transition probability from the bound
state to the scattering state is%
\begin{equation}
1-T_{\pm \pm }-T_{+-}=2U_{0}^{2}p(1-p-U_{0}^{2}p),
\end{equation}%
which is always positive for small nonzero $U_{0}$. This indicates that the
collision between a photon and a polariton can induce the photon emission
from the polariton.

We employ the numerical simulation for verification and demonstration of our
analysis. We compute the time evolution of an initial bound state by taking
a rectangular approximation to a delta function.%
\begin{equation}
V\left( t\right) =\left\{
\begin{array}{cc}
\frac{U_{0}}{w}\left\vert e\right\rangle \left\langle e\right\vert , &
w>t-\tau >0 \\
0, & \text{otherwise}%
\end{array}%
\right. .
\end{equation}

\begin{figure}[tbp]
\centering
\includegraphics[ bb=111 219 378 569, width=0.4\textwidth, clip]{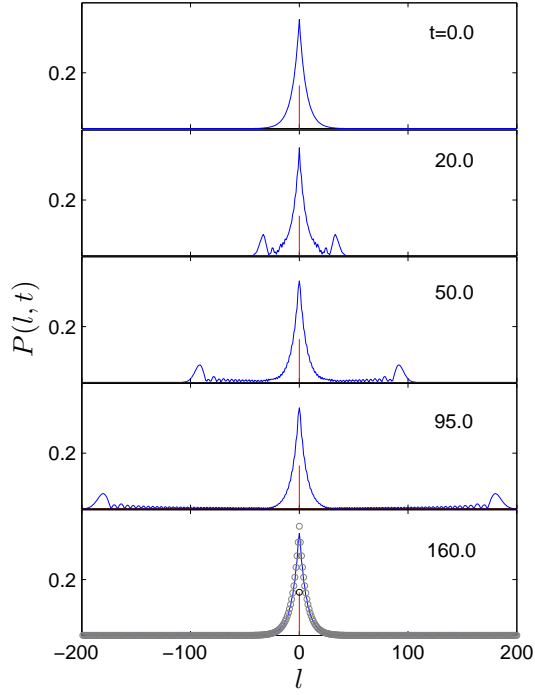}
\caption{(Color online) Time evolution of the initial bound state $%
\left\vert \Phi \left( 0\right) \right\rangle =\left\vert \protect\phi %
^{-}\right\rangle $ in Eq. (\protect\ref{bound+/-}) driven by the
rectangular-pulsed potential. The magnitude distributions of the evolved
wave function $\protect\sqrt{P\left( l,t\right) }$\ for several instants are
obtained as converging results for $\protect\lambda =0.8$, $U_{0}=2$, and $%
w=2\times 10^{-5}\protect\kappa ^{-1}$. The red line indicates the
probability of the $e$th site. The circle(black) represents the initial
profile of the bound state at site $e$ and the circles(gray) represents the
initial profile of the bound state as comparison to the profile of the final
state. It shows that there is a particle probability spreading out from the
center to the infinity, and the final bound state has almost the same shape
as the initial one but less probability. This indicates that a kicked
potential can induce the transition from the bound states to the scattering
states.}
\label{fig4}
\end{figure}

For fixed $U_{0}$, we carry the calculation for different values of $w$. It
is found that the result becomes stable as $w$\ decreases. The convergent
data are adopted as an approximate numerical result. The evolution of an
initial bound state under the central potential pulse is computed as well.
The magnitude distribution of the evolved wave function $\sqrt{P\left(
l,t\right) }=\left\vert \langle l\left\vert \Phi \left( t\right)
\right\rangle \right\vert $ is plotted in Fig. \ref{fig4}. Here the propose
of using $\sqrt{P\left( l,t\right) }$\ rather than the probability $P\left(
l,t\right) $\ is to highlight the escaping wave packets from the center. We
can see that there are two sub-wave packets propagating to the leftmost and
rightmost, and the amplitude of the central bound state is reduced after
this process. It can be predicted that the bound-state probability will keep
decreasing by the successive pulses potential.

The result of this section cannot be regarded as sufficient proof of the
occurrence of the stimulated photon emission from a polariton. Nevertheless,
it shows that there is a high possibility that such a process can happen. In
the following section, we will investigate this phenomenon by numerical
simulation.

\begin{figure*}[tbp]
\includegraphics[ bb=62 624 224 789, width=0.32\textwidth, clip]{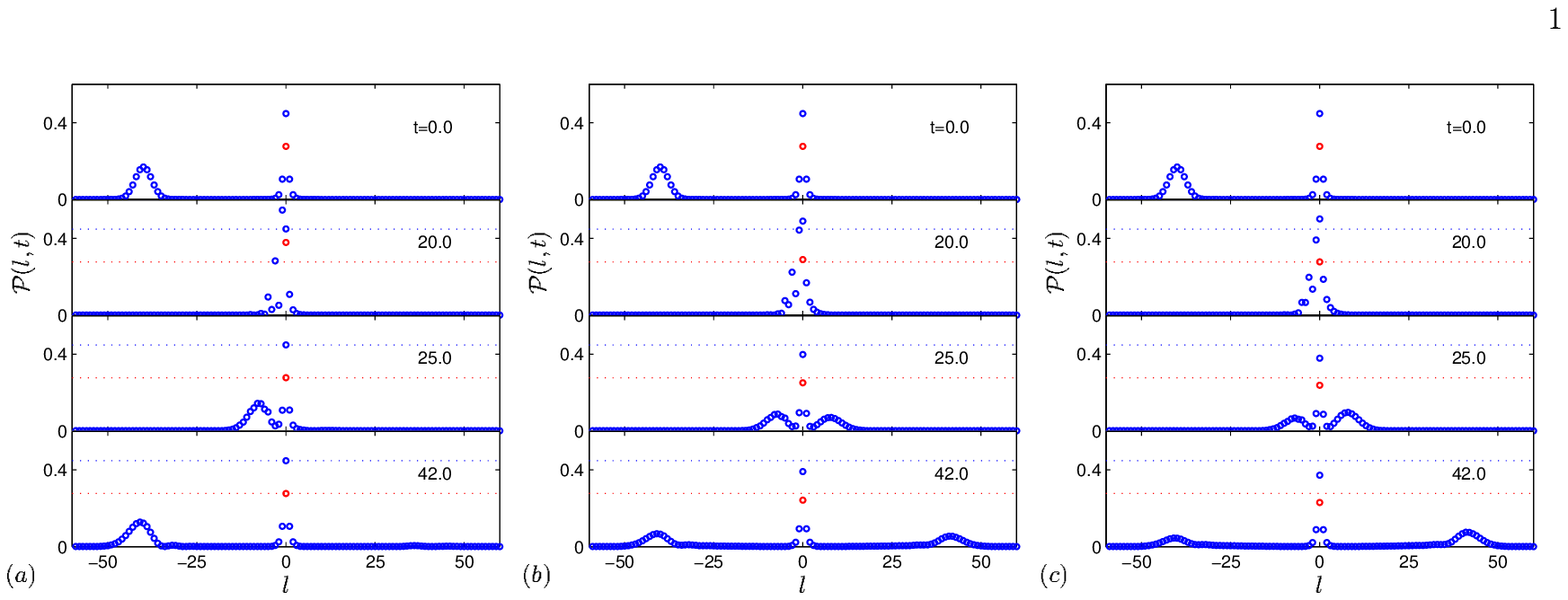} %
\includegraphics[ bb=227 624 389 789, width=0.32\textwidth, clip]{5.eps} %
\includegraphics[ bb=393 624 555 789, width=0.32\textwidth, clip]{5.eps}
\caption{(Color online) Collision process between an incident photon wave
packet and a polariton. The probability distributions $\mathcal{P}\left(
l,t\right) $\ for several instants are obtained by the time evolution\ under
the systems of (a) the Hamiltonian $H_{\text{eq}}$\ with $\protect\lambda =2$%
, $U=0$ and (b) $U=10$, (c) the original Hamiltonian $H$ with $\protect%
\lambda =2$ (or equivalently, $H_{\text{eq}}$\ with $U=\infty $). The
incident wave packet has $k_{0}=\protect\pi /2$\ and $\protect\alpha =0.3$.
The blue (red) dotted line indicates the initial probability of the $0$th ($e $%
th) site as comparison. It shows that the probability of the scattering
photon is conserved for the non-interacting case with $U=0$, but not
conserved in the presence of nonlinearity in $H$. The result demonstrates
the occurrence of stimulated photon emission from a polariton. Moreover, it
is observed that the incident wave packet is totally reflected from the
center in the case (a), but transmitted in the aid of the polariton. The
very close similarity between (b) and (c) indicates that equivalence between
Hubbard and the cavity-atom models. In both cases, the scattered and emitted
photons are still local, keeping the similar shape of the incident one.}
\label{fig5}
\end{figure*}

\begin{figure}[tbp]
\centering
\includegraphics[ bb=46 232 435 545, width=0.45\textwidth, clip]{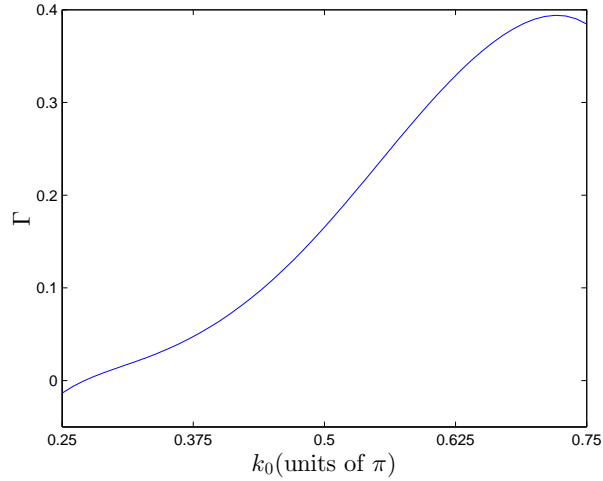}
\caption{(Color online) Emission probability from\ polariton sized $l_{0}=9$%
,\ stimulated by the photon wave packet with $\protect\alpha =0.3$\ and
different $k_{0}$, for the system with $\protect\lambda =2$. It shows that
the transition probability can reach $0.4$ at $k_{0}\approx 0.73\protect\pi $%
.}
\label{fig6}
\end{figure}

\subsection*{Numerical simulation}

In principle, one can explore the problem by solving the coupled equations (%
\ref{coupled eqs}) numerically. The truncation approximation is necessary
since a numerous number of equations are involved. However, we can take an
alternative way for truncation approximation, which is more efficiency for a
discrete system. We can solve the Schrodinger Eq. (\ref{seq}) in finite real
space by computing the time evolution of the initial state

\begin{equation}
\left\vert \Phi \left( 0\right) \right\rangle =\left\vert \varphi
\right\rangle \left\vert \phi ^{-}\right\rangle ,  \label{0state}
\end{equation}%
where $\left\vert \varphi \right\rangle $\ denotes local photonic state
which is separated from polariton $\left\vert \phi ^{-}\right\rangle $ in
real space. The following analysis is also available for the state $%
\left\vert \phi ^{+}\right\rangle $. At time $t$, the evolved state is%
\begin{eqnarray}
\left\vert \Phi \left( t\right) \right\rangle  &=&e^{-iHt}\left\vert \Phi
\left( 0\right) \right\rangle  \\
&=&\sum_{k,k^{\prime }}d_{kk^{\prime }}\left\vert \phi ^{k}\right\rangle
\left\vert \phi ^{k^{\prime }}\right\rangle +\sum_{k,\mu =\pm }a_{k}^{\mu
}\left\vert \phi ^{k}\right\rangle \left\vert \phi ^{\mu }\right\rangle
+\left\vert \xi \right\rangle ,  \notag
\end{eqnarray}%
where $\left\vert \xi \right\rangle $\ denotes two-excitation polaritonic
state. We consider the local photonic state $\left\vert \varphi
\right\rangle $\ as a Gaussian wave packet with momentum $k_{0}$ and initial
center $N_{A}$, which has the form%
\begin{equation}
\left\vert \varphi \left( N_{A},k_{0}\right) \right\rangle =\frac{1}{\sqrt{%
\Omega _{0}}}\sum_{l}e^{-\frac{^{\alpha ^{2}}}{2}%
(l-N_{A})^{2}}e^{ik_{0}l}a_{l}^{\dagger }\left\vert 0\right\rangle
\end{equation}%
where $\Omega _{0}=\underset{l}{\sum }e^{-\alpha ^{2}(l-N_{A})^{2}}$ is the
normalization factor and the half-width of the wave packet is $2\sqrt{\ln 2}%
/\alpha $. We take $2\sqrt{\ln 2}/\alpha \ll \left\vert N_{A}\right\vert $
to ensure the two particles being well separated initially. The evolved wave
function $\left\vert \Phi \left( t\right) \right\rangle $ is computed by
exact numerical diagonalization.

The probability distribution%
\begin{equation}
\mathcal{P}\left( l,t\right) =\left\langle \Phi \left( t\right) \right\vert
a_{l}^{\dag }a_{l}\left\vert \Phi \left( t\right) \right\rangle ,
\end{equation}%
is plotted in Fig. \ref{fig5} to show the profile of the evolved wave
function. One can notice that in the photon-polariton collision process, the
probability of the polariton is not conserved. This result has implications
in two aspects: First, we achieve a better understanding of the occurrence
of stimulated photon emission from a polariton. We find that the scattered
and emitted photons are still local. This is crucial for the multi-polariton
system, since the outcome photons can stimulate the photon emission of
another polariton with high probability. Second, it provides evidence to
support the equivalence between $H_{\text{eq}}$ with large $U$ and the
original $H$.

The above result is for an incident wave packet with $k_{0}=\pi /2$.\ We are
interested in the dependence of emission probability on the central momentum
$k_{0}$ of the incident wave packet. The probability of the survival
polaritons can be measured approximately by the photon probability within
the region of the initial polariton resides in, i.e.,%
\begin{equation}
P_{\mathrm{res}}\left( t\right) =\sum_{\left\vert l\right\vert =e,0}^{l_{0}}%
\mathcal{P}\left( l,t\right) ,
\end{equation}%
where $l_{0}$\ denotes the extent of the polariton. Obviously, $P_{\mathrm{%
res}}\left( t\right) $ contains the probabilities of the residual\ polariton
and the free photons within $\left[ -l_{0},l_{0}\right] $. For infinite
chain system, $1-P_{\mathrm{res}}\left( \infty \right) $\ equals to the
photon emission probability $\Gamma $. In the numerical simulation, the
system is finite, we take $\Gamma =1-$Min$\left[ P_{\mathrm{res}}\left(
t\right) \right] $\ within a finite time interval in order to avoid the
error from the reflected photons. Results of $\Gamma $ as function of $k_{0}$%
\ presented in Fig. \ref{fig6}, show that the maximal photon emission
probability reaches $0.4$ at $k_{0}\approx 0.73\pi $. We can see that the
stimulated transition is significant, which indicates that a polariton is
fragile against an incident photon.

\begin{figure}[tbp]
\centering
\includegraphics[ bb=123 647 306 790, width=0.36\textwidth, clip]{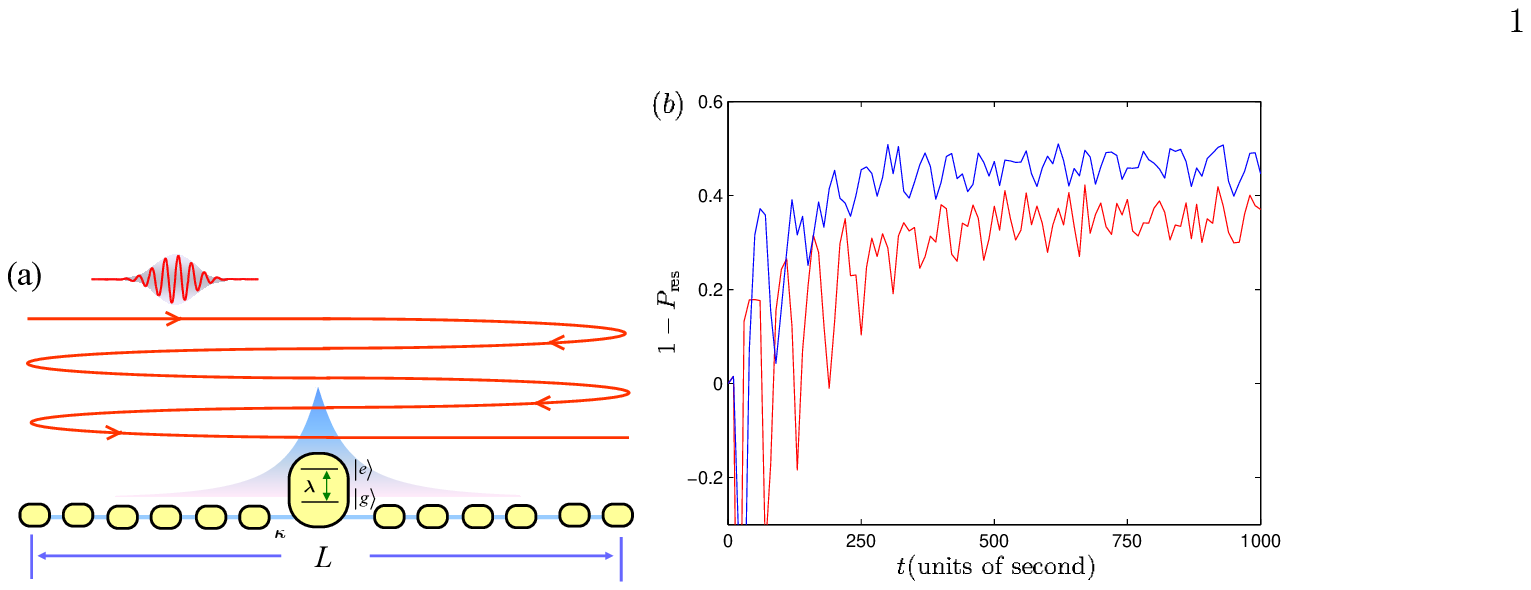} %
\includegraphics[ bb=311 647 492 790, width=0.36\textwidth, clip]{7.eps}
\caption{(Color online) (a) Schematic illustration for the scattering
process of a moving wave packet and a stationary polariton at center of
finite chain. (b) Plots of $1-P_{\mathrm{res}}\left( t\right) $ for the
cases with $\ l_{0}=9$,\ $\protect\alpha =0.3$, $\protect\lambda =2$, $L=120$%
, $k_{0}=3\protect\pi /4$ (blue), $\protect\pi /2$ (red). One can see that
the probability converges to a nonzero constant at long-time scale.}
\label{fig7}
\end{figure}

Now we explore a system with a portion of cavities with doped atom. For a
well prepared insulating phase, which is formed by many independent
polaritons, decreasing $\lambda $\ can lead to the delocalization of the
photons. The above analysis offers an alternative probability: external
radiation can trigger a sudden change of the state. After the collision of
an incident photon and the first polariton, the scattered and emitted
photons can further stimulate other polaritons. In order to mimic such a
chain reaction, we study the multi-collision process\ by computing the time
evolution of the two-particle system\ in a long time scale. We consider a
finite system, in which the scattered and emitted photons are reflected due
to the open boundary condition. It can simulate the repeating collision
process, resulting in the continuous probability decay of polaritons.
Results of our numerical simulations of $1-P_{\mathrm{res}}\left( t\right) $
is presented in Fig. \ref{fig7}(b). It appears that the local average of $P_{%
\mathrm{res}}\left( t\right) $\ continuously decays at beginning as
predicted and then converges to a nonzero constant. As pointed above, $P_{%
\mathrm{res}}\left( t\right) $\ may contain photon probability, leading to $%
P_{\mathrm{res}}\left( t\right) >1$.\ However, the local maxima of $1-P_{%
\mathrm{res}}\left( t\right) $\ can measure the stimulated transition
approximately.

\begin{figure}[tbp]
\centering
\includegraphics[ bb=107 293 404 560, width=0.4\textwidth, clip]{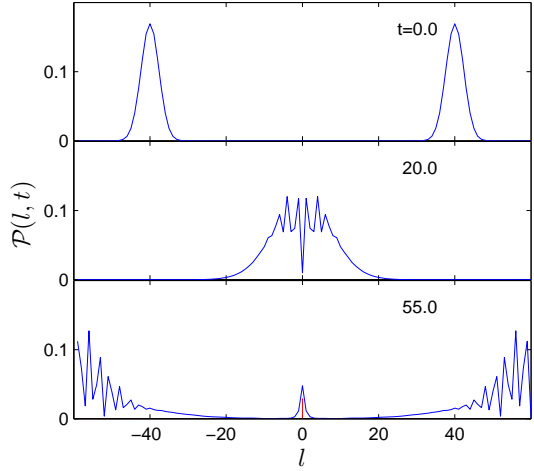}
\caption{(Color online) Collision process between two incident photon wave
packets from leftmost and rightmost, respectively. The probability
distributions $\mathcal{P}\left( l,t\right) $\ for several instants are
obtained by the time evolution\ under the system of the original Hamiltonian
$H$ with $\protect\lambda =2$ (or equivalently, $H_{\text{eq}}$\ with $%
U=\infty $). The red line indicates the probability of the $e$th site. It
shows that the probability of the scattering photon is not conserved in the
presence of nonlinearity in $H$. It demonstrates the polariton can be
created by the collision of two free photons. }
\label{fig8}
\end{figure}

We presume that a polariton should be washed out by successive collision.
However, numerical result shows that the residual polariton probability does
not tend to zero after a long time. There are two main reasons: First, as
time goes on, any wave packets will spread, reducing the impact of photons
on the polariton. Second, the inverse process of photon emission should be
considered, in which two colliding photons can create polaritons. To
demonstrate such a process, we compute the corresponding simulation. In this
process, according to Eq. \ref{0state}, the initial state can be expressed as%
\begin{equation}
\left\vert \Phi \left( 0\right) \right\rangle =\left\vert \varphi \left(
N_{A},\pi /3\right) \right\rangle \left\vert \varphi \left( -N_{A},-\pi
/3\right) \right\rangle ,
\end{equation}%
which implies that there are only two symmetry Gaussian wave packets at the
beginning. At time $t$, the evolved state is%
\begin{eqnarray}
\left\vert \Phi \left( t\right) \right\rangle &=&e^{-iHt}\left\vert \Phi
\left( 0\right) \right\rangle \\
&=&\sum_{k,k^{\prime }}d_{kk^{\prime }}\left\vert \phi ^{k}\right\rangle
\left\vert \phi ^{k^{\prime }}\right\rangle +\sum_{k,\mu =\pm }a_{k}^{\mu
}\left\vert \phi ^{k}\right\rangle \left\vert \phi ^{\mu }\right\rangle
+\left\vert \xi \right\rangle  \notag
\end{eqnarray}%
where $\left\vert \xi \right\rangle $\ denotes the two-excitation
polaritonic state. The probability distribution $\mathcal{P}\left(
l,t\right) $ at several typical instants is plotted in Fig. \ref{fig8}. One
can see that in the photon-photon collision process, the probability of
photons is not conserved as well, which indicates that a polariton can be
created when two photons meet at the $0$-th cavity. This shows that a
polariton can be generated by two-photon Raman scattering.\ As a summary of
numerical results, we conclude that a polariton cannot completely transmit
to a photon by the collision from a single photon, and inversely, a photon
cannot completely transmit to a polariton by the collision from a single
photon. The essential reason is the energy conservation: two-photon energy
cannot match that of one photon plus one polariton, i.e.,%
\begin{equation}
-2\kappa \cos k-2\kappa \cos k^{\prime }\neq -2\kappa \cos k^{\prime \prime
}+\varepsilon _{\pm }.
\end{equation}

This feature can also be employed to realize all-optical control of photon
storage. One main task of quantum information science is to find physical
implementations in which a flying qubit can be stopped to store or process
quantum information. It has been shown that a flying qubit can be stopped
and stored as a collective polariton by tuning the cavity-atom coupling
strength adiabatically \cite{zhou2}. In the present cavity QED system, a
single-photon wave packet can be a flying qubit, while a polariton can be
regarded as a stopped photon, or a stationary qubit. Our result indicates
that a single-photon wave packet, or a train of separated wave packets
cannot excite a polariton if the atom is\ in ground state at the beginning.
Then any incident single photons from one side cannot create a polariton
solely, leaving the atom in the ground state. This can be expressed as
equation
\begin{equation}
\left\langle \phi ^{\pm }\right\vert e^{-iHt}\prod_{i}^{n}\left\vert \varphi
\left( N_{i},k_{0}\right) \right\rangle =0,
\end{equation}%
where $N_{i}<0$, $k_{0}\in \left( 0,\pi \right) $\ and $\left\vert
N_{i+1}-N_{i}\right\vert \gg 2\sqrt{\ln 2}/\alpha $, i.e., all the $n$\ wave
packets incident from left and the neighboring wave packets are well
separated. In contrast a photon can be stopped at the polariton with the aid
of single photons train from the opposite side. This can be expressed as
equation\textbf{\ }%
\begin{equation}
\left\langle \phi ^{\pm }\right\vert e^{-iHt}\left\vert \varphi \left(
\left\vert N_{0}\right\vert ,-k_{0}\right) \right\rangle
\prod_{i}^{n}\left\vert \varphi \left( N_{i},k_{0}\right) \right\rangle \neq
0,
\end{equation}%
i.e, the atom partially absorbs a photon to form a polariton. The processes
expressed by two above Eqs. are schematically illustrated in Fig. \ref{fig3}%
(b)\textbf{.}

\section*{Discussion}

In this paper, the scattering problem of photon and polariton in a
one-dimensional coupled-cavity system has been theoretically investigated.
The analysis shows that, a photon can stimulate the photon emission from a
polariton, which suggests that the insulating phase is fragile against the
external radiation for a system with a lower density of doped cavity. This
result can have some applications in practice. For example, this provides a
way to induce the amplification of the photon population in a
multi-polariton system as a photon amplifier. On the other hand, we also
find that two-photon Raman transition can occur in this cavity QED system,
i.e., a stationary single-excitation polariton can be generated by
three-body, two photons and atom, collision. This phenomenon can be used to
design a scheme to stop and store a single photon. Although this two
photon-polariton transitions is probabilistic, it reveals the peculiar
features of two-excitation dynamics, which significantly differs from a
single-particle scattering problem and opens a possibility to achieve
all-optical control of a single photon. The underlying physics can be
understood as the effective interaction of two photons arising from the
nonlinearity in the doped cavity. These photon emission and absorption
processes is an exclusive signature of correlated photons and could be
applied to the quantum and optical device design.

\section*{Methods}

\subsection*{The exact eigenstates with $\mathcal{N}=1$}

In this section, we present the exact eigenstates with $\mathcal{N}=1$ for
the Hamiltonian $H$. The Hamiltonian has parity symmetry $\left[ P,H\right]
=0$, where $Pa_{l}P^{-1}=a_{-l}$. The odd-parity eigenstates can be obtained
directly, which is%
\begin{eqnarray}
\left\vert \varphi ^{k}\right\rangle &=&\frac{1}{\sqrt{\Omega _{k}}}%
\sum\limits_{l\neq 0}\left( \sin kl\right) a_{l}^{\dagger }\left\vert
g\right\rangle \left\vert 0\right\rangle \\
&=&\frac{\sqrt{2N}}{2i\sqrt{\Omega _{k}}}\left( a_{k}^{\dagger
}-a_{-k}^{\dagger }\right) \left\vert g\right\rangle \left\vert
0\right\rangle  \notag
\end{eqnarray}%
with eigen energy $\varepsilon _{k}=-2\kappa \cos k$, where $\Omega _{k}$\
is the normalization factor and $a_{k}^{\dagger }$\ is the photon operator
in $k$ space, i.e.,%
\begin{eqnarray}
a_{k}^{\dag } &=&\frac{1}{\sqrt{2N}}\sum_{\left\vert l\right\vert
=0}^{N}e^{ikl}a_{l}^{\dag }, \\
a_{l}^{\dag } &=&\frac{1}{\sqrt{2N}}\sum_{k}e^{-ikl}a_{k}^{\dag }.
\end{eqnarray}%
The solutions $\left\vert \phi ^{k}\right\rangle $\ with even parity are two
folds:

(i) For real $k$, the eigenstates has the form%
\begin{equation}
\left\vert \phi ^{k}\right\rangle =g_{k}\left\vert e\right\rangle \left\vert
0\right\rangle +f_{k}a_{0}^{\dagger }\left\vert g\right\rangle \left\vert
0\right\rangle +\sum\limits_{l\neq 0}\left( A_{k}e^{ik\left\vert
l\right\vert }a_{l}^{\dag }+B_{k}e^{-ik\left\vert l\right\vert }a_{l}^{\dag
}\right) \left\vert g\right\rangle \left\vert 0\right\rangle ,
\end{equation}%
where%
\begin{equation}
\left\vert 0\right\rangle =\prod_{\left\vert l\right\vert =0}\left\vert
0\right\rangle _{l},a_{l}\left\vert 0\right\rangle _{l}=0.
\end{equation}%
Submitting $\left\vert \phi ^{k}\right\rangle $\ to the Schrodinger equation%
\begin{equation}
H\left\vert \phi ^{k}\right\rangle =\epsilon _{k}\left\vert \phi
^{k}\right\rangle ,
\end{equation}%
we get the equations for coefficients $g_{k}$, $f_{k}$, $A_{k}$, and $B_{k}$,%
\begin{eqnarray}
&&\epsilon _{k}=-\kappa \left( e^{ik}+e^{-ik}\right) , \\
&&\epsilon _{k}\left( A_{k}e^{ik}+B_{k}e^{-ik}\right) =-\kappa \left(
A_{k}e^{2ik}+B_{k}e^{-2ik}+f_{k}\right) , \\
&&\epsilon _{k}f_{k}=-2\kappa \left( A_{k}e^{ik}+B_{k}e^{-ik}\right)
+\lambda g_{k}, \\
&&\epsilon _{k}g_{k}=\lambda f_{k}.
\end{eqnarray}%
The eigenstates $\left\vert \phi ^{k}\right\rangle $\ are two folds:

(i) For real $k$, a straightforward derivation leads to
\begin{eqnarray}
A_{k} &=&\frac{-g_{k}e^{-ik}}{4i\kappa ^{2}\lambda \sin k}\left[ 2\kappa
^{2}\epsilon _{k}+\left( \lambda ^{2}-\epsilon _{k}^{2}\right) \left(
\varepsilon _{k}+\kappa e^{-ik}\right) \right] \\
&=&\frac{-g_{k}}{4i\kappa \lambda \sin k}\left[ 2\kappa \epsilon
_{k}e^{-ik}-\left( \lambda ^{2}-\epsilon _{k}^{2}\right) \right] ,  \notag \\
B_{k} &=&\frac{g_{k}e^{ik}}{4i\kappa ^{2}\lambda \sin k}\left[ 2\kappa
^{2}\epsilon _{k}+\left( \lambda ^{2}-\epsilon _{k}^{2}\right) \left(
\varepsilon _{k}+\kappa e^{ik}\right) \right] \\
&=&\frac{g_{k}}{4i\kappa \lambda \sin k}\left[ 2\kappa \epsilon
_{k}e^{ik}-\left( \lambda ^{2}-\epsilon _{k}^{2}\right) \right] ,  \notag \\
f_{k} &=&\frac{\epsilon _{k}g_{k}}{\lambda }, \\
\epsilon _{k} &=&-2\kappa \cos k.
\end{eqnarray}%
Then we have%
\begin{eqnarray}
\left\vert \phi ^{k}\right\rangle &=&\frac{1}{\sqrt{\Lambda _{k}}}%
\{\left\vert e\right\rangle \left\vert 0\right\rangle +\frac{\epsilon _{k}}{%
\lambda }a_{0}^{\dagger }\left\vert g\right\rangle \left\vert 0\right\rangle
+\frac{1}{4i\kappa \lambda \sin k}\sum\limits_{l\neq 0}\varsigma _{\pm
}e^{\pm ik\left\vert l\right\vert }a_{l}^{\dag }\left\vert g\right\rangle
\left\vert 0\right\rangle \}, \\
\varsigma _{\pm } &=&\pm \left[ \left( \lambda ^{2}-\epsilon _{k}^{2}\right)
2\kappa \epsilon _{k}e^{\mp ik}\mp \left( \lambda ^{2}-\epsilon
_{k}^{2}\right) \right]
\end{eqnarray}%
where $\Lambda _{k}$\ is the normalization factor, and $\epsilon
_{k}=\varepsilon _{k}=-2\kappa \cos k$. These are extended states.

(ii) There are two eigenstates with complex $k$ which can be seen as two
bound states. The boundary condition%
\begin{equation}
\langle l\left\vert \phi ^{k}\right\rangle =0\text{, for }l\rightarrow \pm
\infty ,
\end{equation}%
and real $\epsilon _{k}$\ require%
\begin{equation}
A_{k}=0,k=i\beta \text{ or }\pi +i\beta
\end{equation}%
with real $\beta >0$. A straightforward derivation leads to

\begin{eqnarray}
B_{k} &=&f_{k}, \\
\lambda ^{2} &=&\kappa ^{2}\left( e^{-2\beta }-e^{2\beta }\right) , \\
\varepsilon _{\pm } &=&\pm 2\kappa \cosh \beta .
\end{eqnarray}

\begin{equation}
e^{2\beta }=\sqrt{\left( \lambda /\sqrt{2}\kappa \right) ^{4}+1}+\left(
\lambda /\sqrt{2}\kappa \right) ^{2}.
\end{equation}%
\bigskip Then we have%
\begin{equation}
\left\vert \phi ^{\pm }\right\rangle =\pm \frac{2\kappa }{\lambda \sqrt{%
\Omega }}\sinh \beta \left\vert e\right\rangle \left\vert 0\right\rangle
+\sum\limits_{\left\vert l\right\vert =0}\frac{\left( \mp 1\right) ^{l}}{%
\sqrt{\Omega }}e^{-\beta l}a_{l}^{\dagger }\left\vert g\right\rangle
\left\vert 0\right\rangle ,
\end{equation}%
where the normalization factor is%
\begin{equation}
\Omega =\left( \frac{2\kappa }{\lambda }\right) ^{2}\sinh ^{2}\beta +\coth
\beta .
\end{equation}%
%
%

\section*{Acknowledgements}

We acknowledge the support of the National Basic Research Program (973
Program) of China under Grant No. 2012CB921900 and CNSF (Grant No. 11374163).

\section*{Author contributions statement}

C.L. did the derivations and edited the manuscript. Z.S. conceived the
project and drafted the manuscript. All authors reviewed the manuscript.

\section*{Additional information}

The authors do not have competing financial interests.

\end{document}